\documentclass[letterpaper]{article} 
\usepackage{aaai2026}  
\usepackage{times}  
\usepackage{helvet}  
\usepackage{courier}  
\usepackage[hyphens]{url}  
\usepackage{graphicx} 
\usepackage{natbib}  
\usepackage{caption} 
\usepackage{algorithm}
\usepackage{algorithmic}

\usepackage{newfloat}
\usepackage{listings}
\DeclareCaptionStyle{ruled}{labelfont=normalfont,labelsep=colon,strut=off} 
\floatstyle{ruled}
\newfloat{listing}{tb}{lst}{}
\floatname{listing}{Listing}

\usepackage{xcolor}
\usepackage{amsfonts,amssymb}
\usepackage{booktabs}
\usepackage{adjustbox}
\usepackage{multirow}
\usepackage{makecell}
\usepackage{diagbox}
\usepackage{subfigure}
\usepackage{marvosym}
\usepackage{pdfpages}
\usepackage{amsmath}

\title{Dynamic Forgetting and Spatio-Temporal Periodic Interest Modeling for Local-Life Service Recommendation}

\author{
    Zhaoyu Hu\thanks{The first two authors contribute equally.},
    Jianyang Wang\footnotemark[1],
    Hao Guo\thanks{ Corresponding author.},
    Yuan Tian,
    Erpeng Xue,
    Xianyang Qi,
    Hongxiang Lin,
    Lei Wang,
    Sheng Chen
}
\affiliations{
    \textsuperscript{\rm}Meituan, Beijing, China\\

    \{huzhaoyu02, wangjianyang, guohao15, tianyuan31, xueerpeng, qixianyang, linhongxiang02, wanglei46, chensheng19\}@meituan.com
}

\usepackage{bibentry}

\begin{document}

\maketitle

\begin{abstract}
In the context of the booming digital economy, recommendation systems, as a key link connecting users and numerous services, face challenges in modeling user behavior sequences on local-life service platforms, including the sparsity of long sequences and strong spatio-temporal dependence. Such challenges can be addressed by drawing an analogy to the forgetting process in human memory. This is because users' responses to recommended content follow the recency effect and the cyclicality of memory. By exploring this, this paper introduces the forgetting curve and proposes \textbf{S}patio-\textbf{T}emporal periodic \textbf{I}nterest \textbf{M}odeling (STIM) with long sequences for local-life service recommendation. STIM integrates three key components: a dynamic masking module based on the forgetting curve, which is used to extract both recent spatiotemporal features and periodic spatiotemporal features; a query-based mixture of experts (MoE) approach that can adaptively activate expert networks under different dynamic masks, enabling the collaborative modeling of time, location, and items; and a hierarchical multi-interest network unit, which captures multi-interest representations by modeling the hierarchical interactions between the shallow and deep semantics of users' recent behaviors. By introducing the STIM method, we conducted online A/B tests and achieved a 1.54\% improvement in gross transaction volume (GTV). In addition, extended offline experiments also showed improvements. STIM has been deployed in a large-scale local-life service recommendation system, serving hundreds of millions of daily active users in core application scenarios.
\end{abstract}


\section{Introduction}

In the rapidly evolving digital economy, recommendation systems are pivotal in linking users with vast services by minimizing decision-making costs and boosting service distribution efficiency through precise matching. The ranking module, a vital component of recommendation systems, directly influences the service list quality users receive \cite{intro1}. Industry research focuses on improving personalization in ranking through user interest mining. User behavior sequences, vital records of interest preferences, encapsulate the whole decision-making journey from browsing to transactions. Deep modeling of these sequences to uncover interest patterns enhances ranking accuracy and holds significant business value in areas like e-commerce and content. With advancements in storage and computing, sequence modeling increasingly trends towards long-sequence implementations. Modeling long user behavior sequences is paramount to capturing the dynamic trends, relevance, and periodicity of user interests, making it a priority for recommendation systems \cite{TWIN, TWIN2, SIM}.

\begin{figure}[t]
\centering
\includegraphics[width=1\columnwidth]{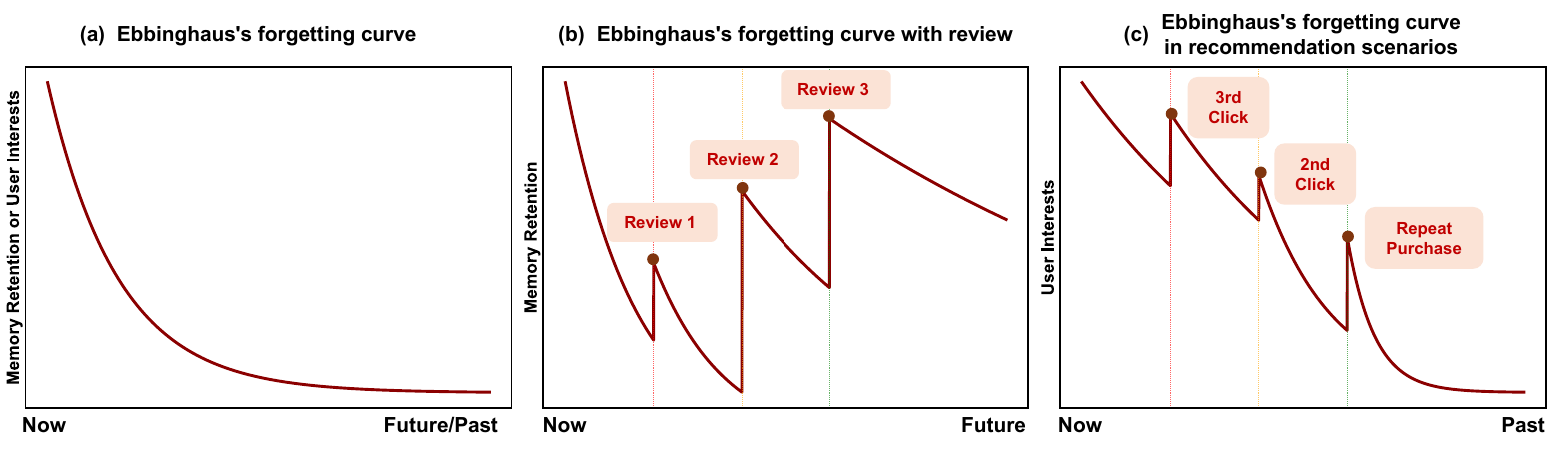} 
\caption{Ebbinghaus's forgetting curve in memory and recommendation scenarios.}
\label{fig1}
\end{figure}

However, user behavior sequence modeling in local-life service platforms poses more complex challenges than in traditional e-commerce or content platforms. Unlike platforms focused on content consumption, a typical local-life service platform, with core businesses (takeaway, shopping, etc.) that inherently integrate temporal-spatial attributes and real-time needs, exhibits unique characteristics in user behavior sequences, including sparsity and strong temporal-spatial dependence. These characteristics present two significant challenges for modeling:

Firstly, long-sequence modeling in local-life scenarios faces a contradiction between effective length and sparsity. While longer sequences can provide more historical information, the distribution of user behaviors on our platform shows occasionality. A user’s last action might have occurred a year ago, or just yesterday. Occasional low-frequency click behaviors may be misjudged as core interests, and the feature learning of effective behaviors may lead to unstable embedding representations due to sparsity, directly impairing the model’s generalization ability. This is particularly challenging for extracting the most critical features of users’ recent behaviors, as the concept of ``recent'' is ambiguous and varies across users.

More critically, local-life services are deeply intertwined with temporal and spatial scenarios, adding complexity to their modeling. User behaviors in these contexts are shaped significantly by both time and location. This strong dependency necessitates two essential capabilities. Firstly, it involves the integration of recurring temporal-spatial patterns, such as identifying weekly dining habits or daily consumption cycles, to capture regular behavioral rhythms. Secondly, it requires four-dimensional collaborative modeling that combines users, time, location, and item types. Only by concurrently modeling these interconnected dimensions can we comprehend the contextual logic behind each behavior and users' latent habits.

\cite{bg1,bg2} have shown that users' responses to recommended content (such as clicks or purchases) follow a psychological principle known as the recency effect, where content closer to the time of request is more likely to attract user interest. In the context of local lifestyle scenarios, users' long-sequence behaviors exhibit patterns similar to the recency effect. Additionally, human memory is characterized by sparsity and spatiotemporal periodicity, which share the same traits as user behavior patterns in recommendation systems. Sparsity refers to the discontinuity in human memory of certain events, such as recalling the last skiing experience and the one before that, which may be separated by a significant time span. Spatiotemporal periodicity, on the other hand, manifests as the recency in memory of repeated events; for example, when recalling brushing teeth in the morning, memory only retains the content of the most recent few brushings. Both the recency effect and memory decay are closely related to a well-known psychological concept, the Ebbinghaus's forgetting curve \cite{IFC, TV}. This gives rise to our motivation: \textbf{to utilize the forgetting curve for modeling the spatiotemporal interests in users' long behavior sequences}.

Figure 1(a) illustrates the forgetting curve, demonstrating that memory strength or user interest gradually diminishes over time. Figure 1(b) shows that memory retention improves with repeated reviews, leading to slower memory decay, as indicated by higher values and a gentler slope in the graph. Conversely, in Figure 1(c), user interest shows an opposite trend—closer repeat interactions to the request moment correlate with higher interest and slower decay, reflected by decreasing values and a steeper slope. This reinforcement effect stabilizes into habits, effectively addressing challenges related to sparse behaviors and spatiotemporal periodicity in local-life scenarios. By simulating memory decay and reinforcement processes, it's possible to filter out noise from low-frequency behaviors, capturing recurring core needs and identifying latent habits precisely. Essentially, the sparsity of long sequences results from low-frequency noise interfering with core interest extraction, which can be mitigated by an adaptive screening method that enhances signals from recent high-frequency behaviors while masking less relevant past weights, thereby reducing sparse data interference in modeling.

To address the aforementioned challenges, this paper proposes the STIM (Spatio-Temporal Periodic Interest Modeling) method, which aims to adaptively extract recent features and spatiotemporal periodic features from user sparse behavior sequences, while performing spatiotemporal joint modeling with adaptive queries. The framework integrates three key components: a dynamic masking module based on the forgetting curve, which adaptively extracts both recent spatiotemporal features and periodic spatiotemporal features; a query-based mixture of experts (MoE) approach, which adaptively activates expert networks for different scenes, enabling the deep extraction of spatiotemporal features and item attributes; and a hierarchical multi-interest network unit, which captures multi-interest representations by modeling the hierarchical interactions between shallow and deep semantics of users' behaviors.

This work has three main contributions:

\begin{itemize}
\item The forgetting curve is combined with dynamic masking for temporal-spatial grouping, enabling the adaptive extraction of recent features and periodic patterns from long user behavior sequences.

\item A hierarchical query expert mechanism and a multi-interest network are proposed, which play the roles of queries in different scenarios and item queries to collaboratively capture users' fine-grained spatio-temporal features and diverse interest representations.

\item The STIM method has been deployed in a large-scale local-life service recommendation system, serving hundreds of millions of daily active users in core application scenarios.
\end{itemize}

\section{Related Work}

\textbf{User Interest, Long-Term Behavior, and Sequential Spatio-Temporal Modeling} have emerged as interconnected core focuses in recommendation systems. Early memory network approaches encoded preferences into fixed representations \cite{10.1145/2487575.2487686, 10.1145/2873055} but struggled with dynamic changes. Deep learning brought advancements: DIN \cite{DIN} introduced target attention, DIEN \cite{DIEN} captured temporal dependencies, SASRec \cite{SASRec} modeled long-range dependencies via self-attention, and Caser \cite{10.1145/3159652.3159656} extracted local patterns with convolutions. However, these faced computational limits with extremely long sequences. Two-stage frameworks (SIM \cite{SIM}, ETA \cite{chen2022efficientlongsequentialuser}, TWIN \cite{TWIN}) addressed scalability but still struggled with full life-cycle behavior capture due to retrieval unit length constraints. Frequency-domain analysis, critical for periodic behaviors in local-life services, saw FEARec \cite{c:2} and BSARec \cite{10.1609/aaai.v38i8.28747} apply Fourier transforms for denoising and bias balancing, respectively. FIM \cite{FIM} advanced this field by introducing a frequency-aware module to capture dynamic periodic patterns, integrating multi-view interest decomposition to distinguish diverse periodicities in user demands—this also enabled adaptation to scenarios like holidays or promotions. Clustering-based methods, such as TWIN-V2 \cite{TWIN2} with hierarchical clustering, compressed long sequences efficiently but, like other frameworks, prioritized scalability over sparse pattern mining and fine-grained interest disentanglement. Sequential spatio-temporal modeling, linking preferences to time and location, has grown vital \cite{chi2023modelingspatiotemporalperiodicitycollaborative, ST-PIL, ijcai2022p490}. Early discrete time embeddings \cite{DIN, 10.1145/3447548.3467178} struggled in real-time streaming due to overfitting. Interest Clock \cite{interestclock} used personalized hour-level encoding but was limited to short-term, coarse-grained features. Scenario-specific models \cite{10.1145/3580305.3599866} addressed spatial patterns but lacked adaptability to diverse temporal granularities. Long-term Interest Clock (LIC) \cite{LIC} extended temporal perception to long-term behaviors via time-gap-aware attention but, like other methods, relied on indirect domain transformations (e.g., graph/frequency) for spatio-temporal modeling, failing to autonomously learn multi-granularity temporal features or account for external events. Unlike existing studies, we simultaneously focus on explicit modeling recent temporal patterns, spatiotemporal periodicity, and spatiotemporal interactions under the condition of sparse long-sequence user behaviors. 

\textbf{Forgetting Curve} has been utilized in recommendation systems to model the temporal decay of user interest or feedback. \cite{IFC} introduced the Interest-Forgetting Curve (IFC) to quantify time's impact on music preferences. The TV recommendation's LMFE model combined traditional forgetting curve with enhancement, optimizing recommendations \cite{TV}. \cite{Nextone} used behavior data and the forgetting curve to track music preferences' fading. PMORS \cite{bg2} applied it to negative feedback, calculating penalty via time-weighted layer, and used Pareto optimization for multi-objective balancing, expanding its role in feedback modeling. However, none of these methods have applied the forgetting curve to the modeling of long user behavior sequences. In this work, we aim to address this gap by STIM.

\section{Methodology}

\subsection{Problem Definition}
In recommendation systems, user historical behavior sequences are denoted as \( B \), where \( m \) represents the length of the sequence, and \( B_1 \) to \( B_m \) are ordered by the chronological order of user behaviors. Other features, including statistical features, dense bucketed features, and user profiles, are collectively denoted as $C$. Ranking tasks can be formulated as:
\begin{equation}
\hat{y} = g\left(f\left(B; \Theta_f\right),C; \Theta_g\right)
\end{equation}
Here, $f$ and $g$ represent the behavior sequence model and the prediction layer model respectively, with learnable parameters $\Theta_f$ and $\Theta_g$. The core objective of this paper is to extract spatio-temporal features from the input sequence through $f$.


\begin{figure}
\centering
\includegraphics[width=1\columnwidth]{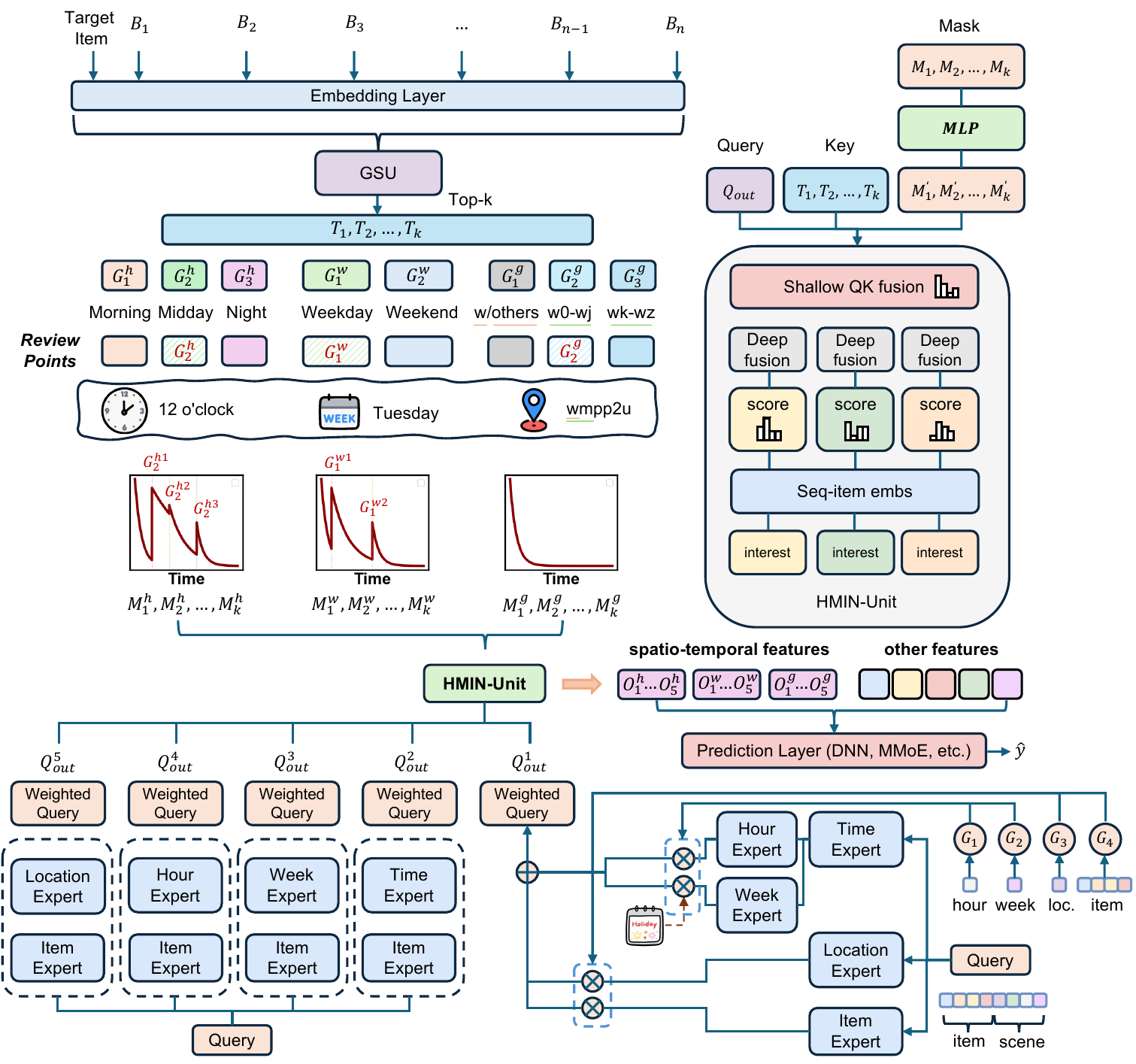} 
\caption{Overall architecture of the spatio-temporal periodic interest model.}
\label{fig2}
\end{figure}

\subsection{Spatiotemporal Dynamic Masking with Forgetting Curve}

Inspired by the SIM (Pi et al. 2020), we first perform compression through General Search Unit (GSU) based on the target item. GSU operation generate a new sequence of length \( k\), denoted as \( T_1, T_2, \ldots, T_k \). The core motivation is leveraging the forgetting curve to introduce temporal decay mechanisms and extract periodic features. The definition of ``review points'' in the forgetting curve is critical. In this work, spatiotemporal data are grouped such that instances falling within the same group as the request moment are treated as review points in the forgetting curve. 

For spatiotemporal grouping, time is divided into hourly and daily groups, both of which exhibit inherent periodic patterns. Regarding spatial grouping, each point of interest is associated with latitude and longitude coordinates, which are transformed into geohash6 representations. One month of platform data was analyzed to calculate the average click frequency per hour and the frequency distribution of the first and second characters in geohash6 for each click location. Based on industry experience, the daily time is partitioned into Morning (3:00–10:00), Midday (11:00–16:00), and Night (17:00–2:00). Weekly time divides into weekdays and weekends. For geohash6\footnote{https://en.wikipedia.org/wiki/Geohash}, the first character ``w'' accounts for the majority (approximately 50\%), while the second character shows a bimodal distribution: ``0-j'' and ``k-z'' each account for half of the total. Three groups are thus created based on the following rules: (1) first character is non-``w''; (2) first character is ``w'' and second character is ``0-j''; (3) first character is ``w'' and second character is ``k-z''. After grouping the input \( T_1,T_2,\ldots,T_k \) by Hour, Week, and geohash6, the results are expressed as \( G_h=\{G_{1h},G_{2h},G_{3h}\} \), \( G_w=\{G_{1w},G_{2w}\} \), and \( G_g=\{G_{1g},G_{2g},G_{3g}\} \), respectively. To determine the spatiotemporal group of a user's query, a hard search is performed based on the request moment. For example, as illustrated in Figure 2, a query submitted on Tuesday at 12:00 from location ``wmpp2u'' falls into groups \( G_{2h} \), \( G_{1w} \), and \( G_{2g} \). All timestamps in these groups are identified as review points.

As shown in Figure 1, the forgetting curve adopts the exponential decay model of the classic Ebbinghaus's forgetting curve, formulated as:
\begin{equation}
R(t) = e^{(-t/S)}
\end{equation}
where \( R(t) \) denotes the memory retention rate; \( S \) is the characteristic time constant of the forgetting curve.

The review effect is adjusted incrementally with the number of reviews, which can be expressed as a piecewise function:
\begin{equation}
R_n =
\begin{cases}
R_{\text{init}} &  n=1, \\
R_{\text{init}} + (R_{\text{final}} - R_{\text{init}}) \times \frac{i-1}{n-1} &  n>1
\end{cases}
\end{equation}
where $i$ denotes the $i$-th review in the $n$-th round of reviews. In this manner, \(R_n\) will be constrained to decay within the range from \(R_{\text{init}}\) to \(R_{\text{final}}\). Additionally, the current decay rate is given by \(D_n = S / (1 + n \times I)\).

Memory retention after review is calculated as:
\begin{equation}
\label{eq:example}
R(t) = R_n e^{(-(t - t_{\text{last}}) \times D_n)}
\end{equation}
where \( R_n \) is the retention rate after the review; \( D_n \) is the decay rate after the review; \( t_{\text{last}} \) is the timestamp of the previous review.

In recommendation scenarios, the trends of the forgetting curve's decay rate and post-review values are opposite to those in traditional scenarios. Specifically, in Equation 3, the memory retention $R_n$ should increase with the number of reviews, and the forgetting rate $D_n$ should slow down. However, in recommendation scenarios, the user interest retention $R_n$ should decrease as the number of reviews increases, and concurrently, the forgetting rate $D_n$ should accelerate. This is because additional review points are progressively further from the user's request time. Based on this, $R_n$ and $D_n$ can be redefined as follows:

\begin{equation}
R_n =
\begin{cases}
R_{\text{final}} &  n=1, \\
R_{\text{final}} - (R_{\text{final}} - R_{\text{init}}) \times \frac{i-1}{n-1} &  n>1
\end{cases}
\end{equation}
The decay rate is modified to \( D_n = (1 + n \times I) / S \). Finally, user interests are calculated by applying Equation 2 initially and switching to Equation 4 after the review point. Based on this, the dynamic mask is derived by mapping time onto the sequence length and performing min-max scaling normalization, which preserves relative differences.
\begin{equation}
M(k) = \sigma(R(t))
\end{equation}

\cite{proof1} has proven that materials of different natures exhibit distinct forgetting curves. In this work, hour, week, and geohash are treated as materials of different natures, thus corresponding to different forgetting curves. For \( M_{h}^1, M_{h}^2, \ldots, M_{h}^k \), \( M_{w}^1, M_{w}^2, \ldots, M_{w}^k \), and \( M_{g}^1, M_{g}^2, \ldots, M_{g}^k \) derived from hour, week, and geohash respectively, new masks \( M_1', M_2', \ldots, M_k' \) for different material types are obtained through a layer of MLP processing. The MLP layer enables simple mapping transformations on the input masks, enhancing waveform diversity and training stability.

To clearly depict the entirety of the forgetting process in recommendation contexts, Figure 3 shows the decay of the forgetting curve from varied angles following equidistant sampling. The curve begins at the user's sequence nearest to the request time. For hour-based grouping, the weight peaks at the moment closest to the request and gradually wanes. When hitting a "review point," the weight increases again but does not return to the original peak, as recent user actions have the greatest impact, succeeded by long-term periodic effects. Notably, near review points, non-group behaviors are also given significant weight due to the necessity of accounting for behavior causality—actions at a given time are influenced by prior adjacent behaviors. In geohash-based grouping, the query doesn’t fit any specific category, adhering to the basic forgetting curve where only recent behavior sequences influence the results. Through this mechanism, this study enables the adaptive extraction of recent user behavior sequences and the capture of long-term periodic spatio-temporal features of users, including the extraction of users' long-term behavioral habits, given that repeated ``reviews'' are more conducive to habit formation.

\begin{figure}
\centering
\includegraphics[width=1\columnwidth]{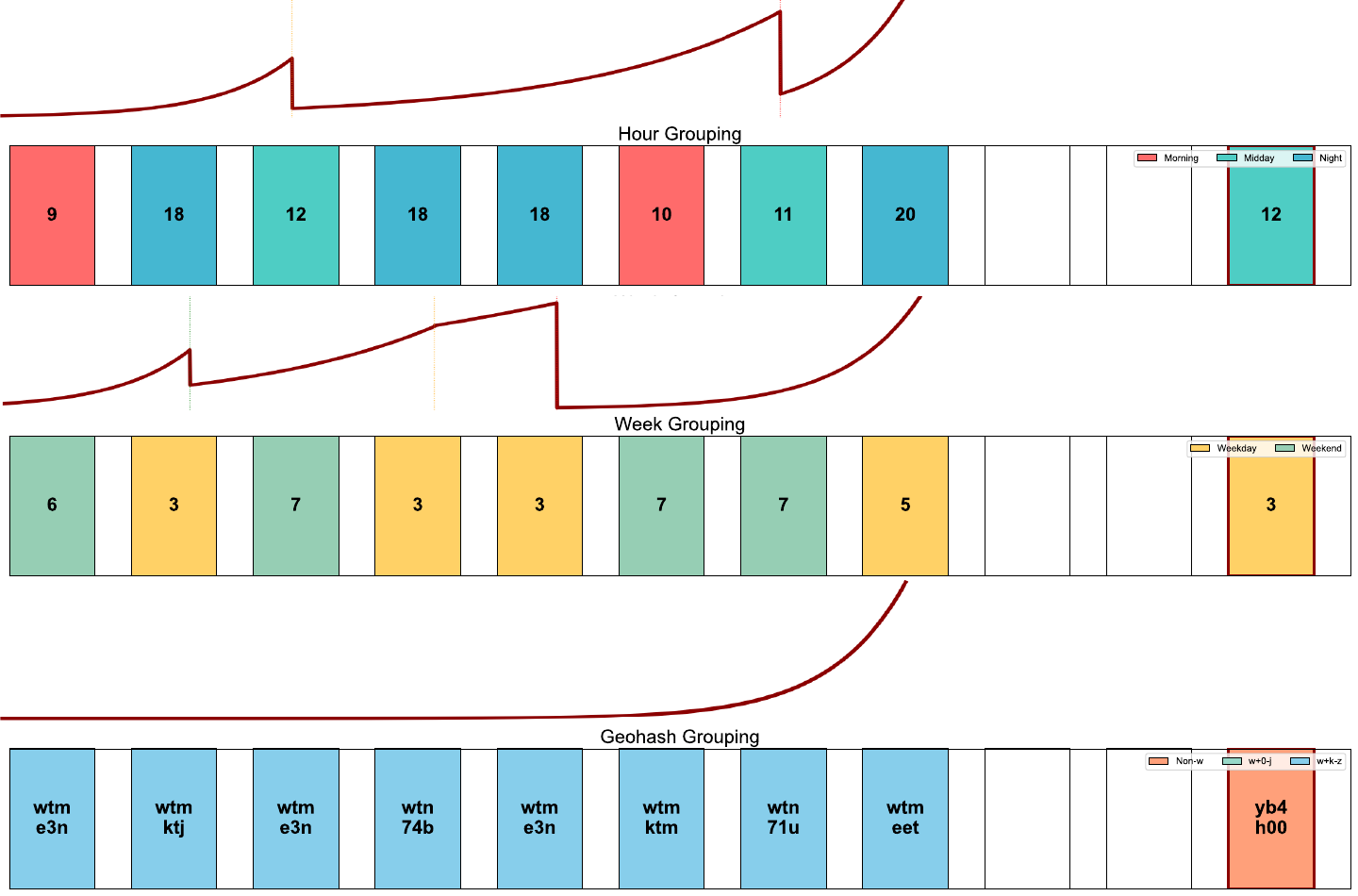} 
\caption{Illustration of forgetting curve decay in recommendation scenarios. The rightmost block depicts the request query, the white blocks indicate padding, and the left blocks represent historical user behavior sequences ordered chronologically from the farthest to the nearest.}
\label{fig3}
\end{figure}

\subsection{Query Weighting via Mixture of Experts}

The dynamic masking method organizes spatio-temporal information into groups and injects spatio-temporal priors into the network. However, the impact of these three groupings: hour, week, and geohash, on final classification cannot be directly learned by the network. To achieve more fine-grained spatio-temporal feature extraction, the network must independently learn to perform joint modeling of time, location, and items based on queries, thereby forming a complementary synergy with the dynamic masking module. Additionally, holiday factors, as special behavioral patterns, are crucial for spatio-temporal modeling, and the model must also adapt to sudden shifts in user preferences driven by holidays.

Building upon this, the paper introduces Query MoE, as shown in Figure 2. The implementation of Query MoE involves first calculating gating weights on the independent spatio-temporal queries, and then applying these weights to the original input queries for joint modeling. Finally, by introducing pairwise combinations of experts, the network can also identify distinctive patterns that are primarily influenced by time and location. If the query time falls on a holiday, an additional holiday factor weight is incorporated into the week expert to enhance day-level feature extraction.

The input \(\mathbf{X}\) is processed by three base experts to generate initial feature representations:
\begin{equation}
h_{\text{Time}}, h_{\text{Loc}}, h_{\text{Item}} = E_{\text{Time}}(\mathbf{X}), E_{\text{Loc}}(\mathbf{X}), E_{\text{Item}}(\mathbf{X})
\end{equation}

The output of the time expert \(h_{\text{Time}}\) is further processed by two sub-experts to capture hour-level and week-level periodic features:
\begin{equation}
h_{\text{Hour}}, h_{\text{Week}} = E_{\text{Hour}}(h_\text{Time}), E_{\text{Week}}(h_\text{Time})
\end{equation}

The hour, week, location, and item within the input \(\mathbf{X}\) can be denoted as \(\mathbf{X}_\text{Hour}\), \(\mathbf{X}_\text{Week}\), \(\mathbf{X}_\text{Loc}\), and \(\mathbf{X}_\text{Item}\) respectively. Gating units \(G_1\) and \(G_2\) can output weights for the hour and week sub-experts, denoted as \(\omega_{\text{Hour}}\) and \(\omega_{\text{Week}}\) respectively. These weights are calculated as:
\begin{equation}
\omega_{\text{Hour}}, \omega_{\text{Week}} = G_1(X_{\text{Hour}}), G_2(X_{\text{Week}})
\end{equation}

The indicator function \(\mathbb{I}_{\text{holiday}}(t)\) is used to strengthen the input of the week expert. It is able to determine whether time $t$ is a holiday, returning a value of 0 or 1. The enhanced weight is:
\begin{equation}
\omega_{\text{Week}}^{'} = \omega_{\text{Week}} + \alpha_{\text{holiday}} \times \mathbb{I}_{\text{holiday}}(t)
\end{equation}
where \(\alpha_{\text{holiday}}\) is the holiday enhancement coefficient. The results are summed to get the intermediate output T:
\begin{equation}
T = \omega_{\text{Hour}} \odot h_{\text{Hour}} + \omega_{\text{Week}}^{'} \odot h_{\text{Week}}
\end{equation}

Gating unit \(G_3\) and \(G_4\) outputs weights for the location expert and item expert, denoted as \(\omega_{\text{Loc}}\) and \(\omega_{\text{Item}}\) respectively:
\begin{equation}
\omega_{\text{Loc}}, \omega_{\text{Item}} = G_3(X_{\text{Loc}}), G_4(X_{\text{Item}})
\end{equation}

These weights are applied to their corresponding feature representations and summed to generate the weighted query:
\begin{equation}
Q^1_{\text{out}} = T + \omega_{\text{Loc}} \odot h_{\text{Loc}} + \omega_{\text{Item}} \odot h_{\text{Item}}
\end{equation}

As noted in \cite{deepfm}, DNNs inherently excel at capturing implicit feature interactions but lack the ability to model explicit cross-feature relationships. Since each expert in our architecture is implemented as a DNN, we draw inspiration from \cite{FM} to introduce explicit interaction mechanisms—specifically, through pairwise combinations (e.g., temporal-item, spatial-item couplings)—to disambiguate behavior patterns dominated by time versus location. This design enables lightweight explicit interactions between dimensions, avoiding feature collapse caused by direct multi-dimensional entanglement. The weighted query outputs generated by these pairwise combinations are denoted as \(Q^2_{\text{out}}\) to \(Q^5_{\text{out}}\), as illustrated in Figure 2.

\subsection{Hierarchical Multi-Interest Network Unit}

The HMIN-Unit is used to extract users' multiple interest features. It decomposes attention computation into two stages: shallow interaction and deep interaction. Shallow interaction captures surface-level semantic associations between Queries and Keys using cosine similarity, forming the basis for deeper modeling. Deep interaction leverages multi-head dedicated networks to process contextual dependencies and generate diverse interest representations.

For a key sequence $K \in \mathbb{R}^{B \times L \times d_k}$, a query vector $Q^i_{out} \in \mathbb{R}^{B \times d_q}$, and a mask $M' \in \mathbb{R}^{B \times L}$, where $B$ is the batch size, $L$ is sequence length, $d_k$ is the dimension of the key vector and $d_q$ is the dimension of the query vector, the output is a concatenated vector $\text{output} \in \mathbb{R}^{B \times (H \cdot d_k)}$, representing $H$ sets of interest derived from $H$ attention heads.

In shallow interaction, the query is expanded to match the key dimensions using a feedforward network, normalized, and compared with the keys through cosine similarity. Dynamic spatio-temporal masks \( M' \) are applied to emphasize and highlight the valid keys.
\begin{equation}
\hat{Q}^i_{out} = \text{L2Norm}\left( \text{DNN}_Q \left( \text{ExpandDim}(Q^i_{out}) \right) \right)
\end{equation}
\begin{equation}
S^i_{\text{masked}} = \hat{Q}^i_{out} \cdot \text{L2Norm}(K)^\top \odot M'
\end{equation}

In deep interaction, each attention head applies a dedicated feedforward network to the masked similarity scores to generate context-sensitive weights.
\begin{equation}
w^h_i = \text{Softmax}(S^i_{\text{masked}})    
\end{equation}
\begin{equation}
O_i = Concat(w_i^{h} \odot \text{DNN}_h(K)) \quad 
\end{equation}
In this context, $i$ and $h$ denote the weighted query index and attention head index, respectively. Eventually, the five outputs obtained separately from hour, week, and geohash are concatenated to form the spatio-temporal features. These are combined with other features $C$ and passed through various prediction heads such as DNN or MMoE to generate the CTR or CTCVR prediction result $\hat{y}$. At this point, the objective can be optimized by calculating the cross-entropy loss between $\hat{y}$ and the true label $y$.
\begin{equation}
\mathcal{L} = -y \log \hat{y} - (1 - y) \log (1 - \hat{y})
\end{equation}

\section{Experiments}
\begin{table}
\setlength{\tabcolsep}{3.7pt} 
\centering
\begin{tabular}{ccccc} 
\toprule
\textbf{Dataset} & \textbf{\#Users} & \textbf{\#Items} & \textbf{MaxLen} & \textbf{Timespan} \\ \midrule
Ele.me & 14 million & 7 million & 50 & 30 \\ 
TRec &  786 million & 162 million & 20000 & 365 \\ 
\bottomrule
\end{tabular}
\caption{Statistics of datasets.}
\end{table}

\begin{table*}[ht]
    \centering
    \begin{adjustbox}{max width=\textwidth}
    \begin{tabular}{ccccccccccc}
    \toprule
    \multirowcell{4}{\textbf{Models}} &
     \multirowcell{4}{\textbf{Recent}} & \multirowcell{4}{\textbf{Temporal}} & \multirowcell{4}{\textbf{Spatial}} & \multicolumn{2}{c}{\textbf{Ele.me}} & \multicolumn{4}{c}{\textbf{TRec}} \\
    \cmidrule(lr){5-6} \cmidrule(lr){7-10}
    & & & & \multicolumn{2}{c}{\textbf{CTR}} & \multicolumn{2}{c}{\textbf{CTR}} & \multicolumn{2}{c}{\textbf{CTCVR}} \\
    \cmidrule(lr){5-6} \cmidrule(lr){7-8} \cmidrule(lr){9-10}
    & & & & \textbf{AUC} & \textbf{GAUC} & \textbf{AUC} & \textbf{GAUC} &
    \textbf{AUC} & \textbf{GAUC} \\
    \midrule
    GSU & & & & 0.5547 & 0.5453 & \underline{0.7662} & 0.6893 & 0.8690 & 0.7488 \\
    \midrule
    DIN & & & & 0.5745 & 0.5516 & 0.7650 & 0.6892 & 0.8682 & 0.7514 \\
    DIEN & & \checkmark & & 0.5950 & 0.5892 & 0.7652 & 0.6888 & 0.8683 & 0.7517 \\
    \midrule
    SIM & \checkmark & \checkmark & & 0.5930 & 0.6208 & 0.7661 & 0.6893 & 0.8690 & 0.7489 \\
    TWIN V2 & \checkmark & & & \underline{0.6285} & \underline{0.6423} & 0.7658 & 0.6903 & \underline{0.8695} & \underline{0.7541} \\
    \midrule
    IC & & \checkmark & & 0.5684 & 0.5695 & 0.7657 & 0.6902 & 0.8692 & 0.7538 \\
    LIC & & \checkmark & & 0.5830 & 0.5794 & 0.7658 & 0.6900 & 0.8693 & 0.7538 \\
    \midrule
    FIM & & \checkmark & \checkmark & 0.6003 & 0.6343 & 0.7657 & \underline{0.6912} & 0.8694 & 0.7535 \\
    ST-PIL & \checkmark & \checkmark & \checkmark & 0.5963 & 0.5930 & 0.7652 & 0.6898 & 0.8671 & 0.7540 \\
    \midrule
    \textbf{STIM} & \checkmark & \checkmark & \checkmark & \textbf{0.6884} & \textbf{0.6918} & \textbf{0.7674} & \textbf{0.6916} & \textbf{0.8710} & \textbf{0.7551} \\
    \bottomrule
    \end{tabular}
    \end{adjustbox}
    \caption{Performance comparison of various models on Ele.me and TRec. Every improvement has been verified as statistically significant through a paired t-test, with the results showing a p-value of 0.05 or less.}
    \label{tab:overall2}
\end{table*}

\subsection{Experimental Settings}

\textbf{Datasets.}  We use the public Ele.me dataset\footnote{https://tianchi.aliyun.com/dataset/131047} and the in-house TRec dataset. Ele.me is a well-known takeaway platform that has collected data spanning 8 days, totaling 146 million samples, with the first 7 days designated for training and the last day for evaluation. The TRec dataset is constructed from real-world, industrial-grade user purchase behavior data sourced from a leading e-commerce platform. This dataset encompasses millions of users, with their behavioral records covering the past two months. We employ time-based splitting: data from days [1, T] is used for training, and day T+1 is used for testing. The statistics of the datasets are shown in Table 1. In the table, MaxLen represents the maximum length of the user behavior sequence, while Timespan indicates the time span of the user behavior sequence.

\textbf{Baselines.} To rigorously evaluate the effectiveness of STIM, we systematically compared it against four categories of state-of-the-art methods: (1) interest mining methods (DIN \cite{DIN}, DIEN \cite{DIEN}); (2) long-sequence modeling methods (SIM \cite{SIM}, TWIN V2 \cite{TWIN2}); (3) temporal interest modeling methods (IC \cite{interestclock}, LIC \cite{LIC}); and (4) spatiotemporal modeling methods (FIM \cite{FIM}, ST-PIL \cite{ST-PIL}). The baseline method only uses GSU for compression.

\textbf{Evaluation Metrics.} We selected the Area Under Curve (AUC) and Group Area Under Curve (G-AUC) for both Click-Through Rate (CTR) and Click-to-Conversion Rate (CTCVR), which are widely adopted in academic and industrial research.

\subsection{Overall Performance}
The overall performance is summarized in Table ~\ref{tab:overall2}. The STIM model proposed in this paper significantly outperforms other models on both the Ele.me dataset and the TRec dataset, achieving remarkable improvements in both CTR and CTCVR metrics. Specifically, on the Ele.me dataset, the AUC and GAUC for CTR are improved by 9.531\% and 7.707\% respectively compared to the sub-optimal model. On the TRec dataset, in terms of CTR, STIM achieves an increase in AUC and GAUC of 0.157\% and 0.058\% respectively over the sub-optimal model. In terms of CTCVR, the improvements are approximately 0.173\% and 0.133\% in AUC and GAUC, respectively.

\subsection{Ablation Study}
The TRec dataset contains both real-world recommendation scenarios and complete CTR and CTCVR labels, making it ideal for conducting the following ablation experiments.

\textbf{Comparison of the foundational functions of forgetting curves.} As illustrated in Figure 4, this paper compares the CTCVR performance of different forgetting curve functions, including exponential, power, and logarithmic functions, ordered from left to right in the figure. Detailed formulas can be found in the supplementary materials. The figure clearly shows that the exponential function achieves optimal and most stable results compared to the other functions, consistent with the findings of \cite{Nextone, bg2}. This provides a solid foundation for its adoption as the form of the forgetting curve function.

\textbf{Effects of hyperparameters on the forgetting curve.} This section performs ablation experiments to investigate the effects of four hyperparameters in the forgetting curve: \(R_\text{init}\), \(R_\text{final}\), $S$, and $I$, as illustrated in Figure 5. The figure reveals that when \(R_\text{init} = 0.4\), \(R_\text{final} = 0.9\), \(S = 20\), and \(I = 2\), although the performance is not the highest in some individual metrics, the overall sum of all metrics reaches the maximum. Additionally, the results indicate that $S$ exerts a more significant impact on the outcomes compared to the other hyperparameters, as it governs the decay rate of the entire curve.

\begin{figure}[t]
\centering
\includegraphics[width=1\columnwidth]{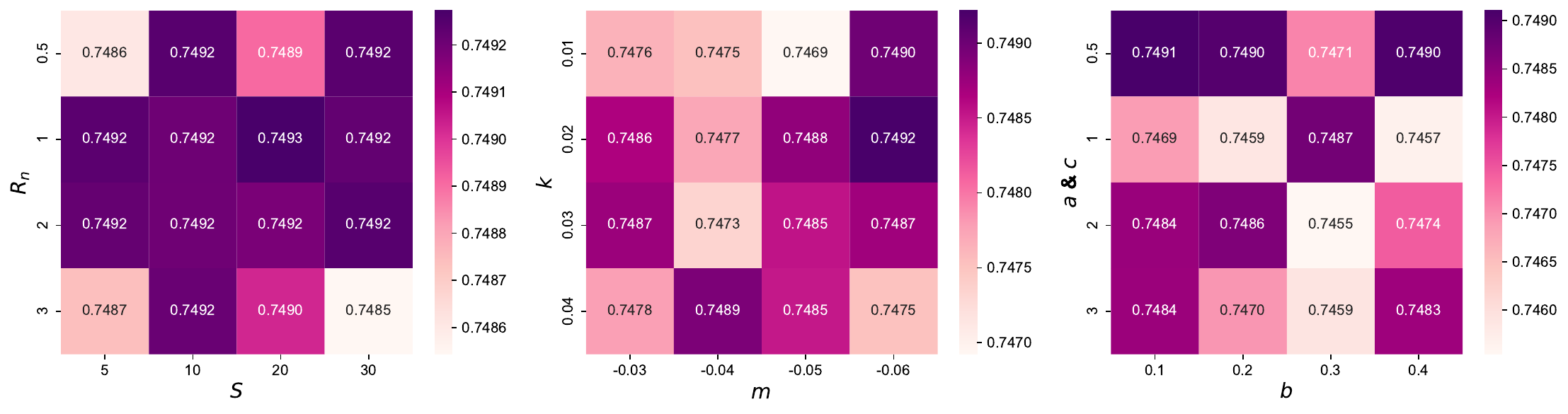} 
\caption{Comparison of the foundational functions of forgetting curves.}
\label{fig4}
\end{figure}

\begin{figure}[t]
\centering
\includegraphics[width=1\columnwidth]{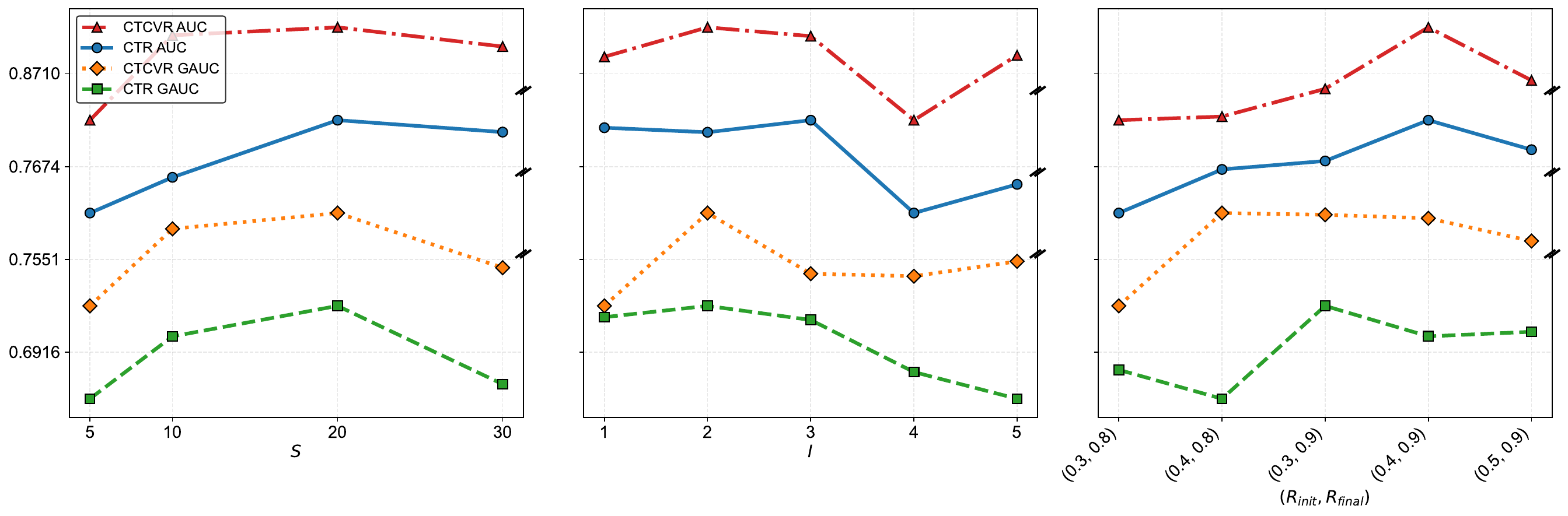} 
\caption{Effects of hyperparameters on the forgetting curve of TRec.}
\label{fig5}
\end{figure}

\begin{figure}[t]
\centering
\includegraphics[width=1\columnwidth]{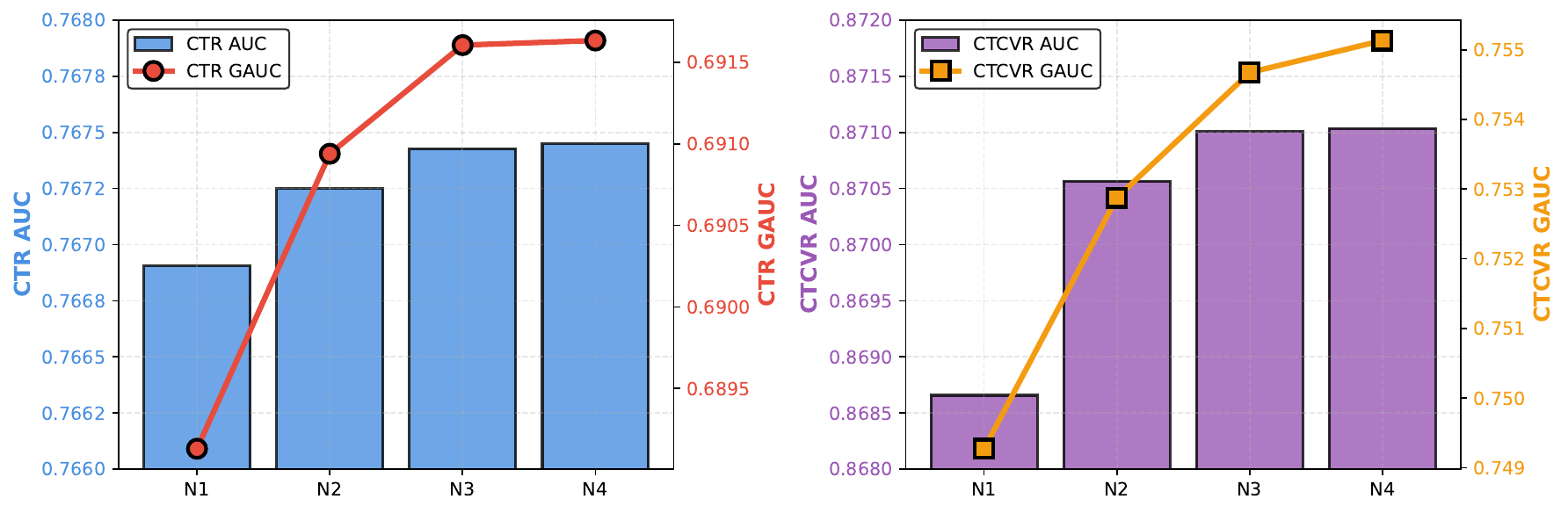} 
\caption{Effects of key components and strategies of spatiotemporal dynamic masking on TRec.}
\label{fig6}
\end{figure}

\begin{figure}[t]
\centering
\includegraphics[width=1\columnwidth]{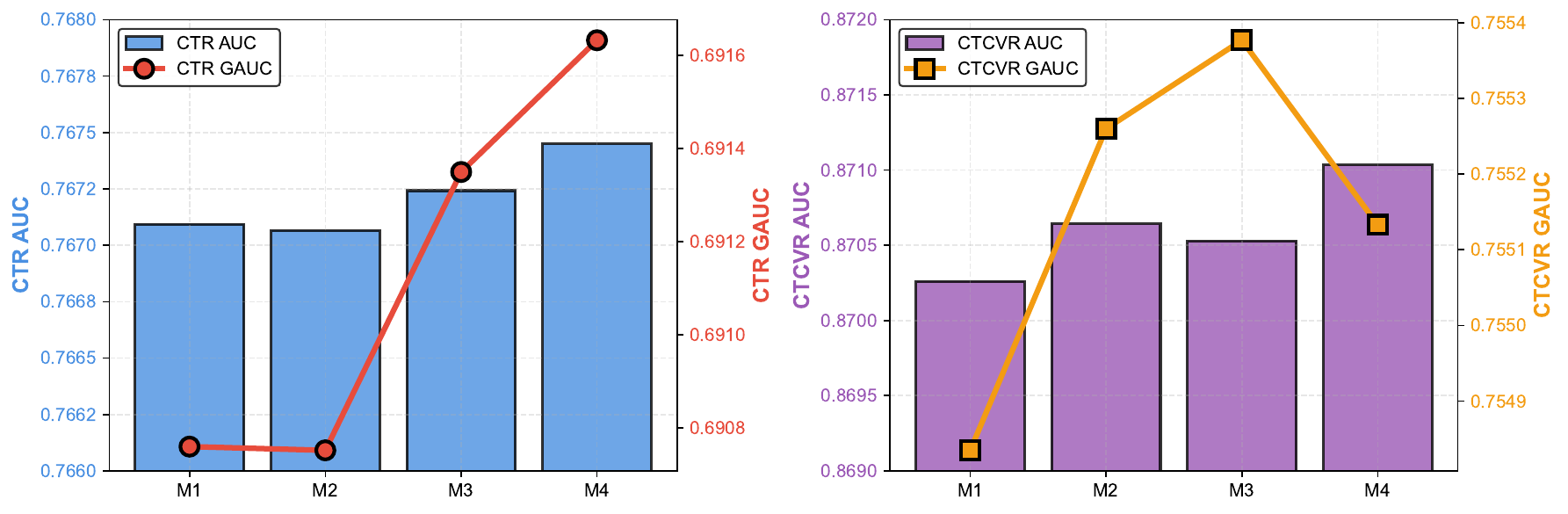} 
\caption{Effects of key components and strategies of Query MoE on TRec.}
\label{fig7}
\end{figure}

\begin{table}
    \centering
    \begin{adjustbox}{max width=\textwidth}
    \begin{tabular}{ccccc}
    \toprule
    \multirowcell{2}{\textbf{Models}} & \multicolumn{2}{c}{\textbf{CTR}} & \multicolumn{2}{c}{\textbf{CTCVR}} \\
    \cmidrule(lr){2-3} \cmidrule(lr){4-5}
    & \textbf{AUC} & \textbf{GAUC} & \textbf{AUC} & \textbf{GAUC} \\
    \midrule
    GSU & 0.7499 & 0.6571 & 0.7828 & 0.7178 \\
    \midrule
    DIN & 0.7480 & 0.6733 & 0.8028 & 0.7148 \\
    DIEN & 0.7496 & 0.6775 & 0.8083 & 0.7163 \\
    \midrule
    SIM & 0.7515 & 0.6771 & 0.8079 & 0.7076 \\
    TWIN V2 & \underline{0.7525} & 0.6806 & 0.8086 & \underline{0.7183} \\
    \midrule
    IC & 0.7515 & 0.6752 & 0.8044 & 0.7031 \\
    LIC & 0.7491 & 0.6790 & 0.8052 & 0.7174 \\
    \midrule
    FIM & 0.7497 & \underline{0.6838} & \underline{0.8167} & 0.7132 \\
    ST-PIL & 0.7461 & 0.6837 & 0.8052 & 0.7110 \\
    \midrule
    \textbf{STIM} & \textbf{0.7553} & \textbf{0.6863} & \textbf{0.8397} & \textbf{0.7365} \\
    \bottomrule
    \end{tabular}
    \end{adjustbox}
    \caption{Cold start performance comparison on TRec.}
    \label{tab:overall3}
\end{table}

\textbf{Effects of various spatiotemporal dynamic masking strategies.} As shown in Figure 6, for the spatiotemporal dynamic masking component, this paper compares four ablation experiments labeled \underline{N1}-\underline{N4}. \underline{N1} directly applies the most classic forgetting curve to the input sequence without incorporating review points. \underline{N2} includes grouping and review points, but sets the review points using identical values instead of based on query hits within the same group. \underline{N3} does not map and transform the dynamic masks output by hour, week, and geohash through a single-layer MLP. \underline{N4} refers to the proposed method. The figure shows that each module and strategy improves performance across the four metrics, with the incorporation of review points offering the most significant enhancement. Additionally, treating the same group as review points rather than identical values also yields notable improvements.

\textbf{Ablation experiments on Query MoE module.} In Figure 7, \underline{M1} refers to directly concatenating the item and scene information at the time of the request as the query. \underline{M2} uses the weighted sum of four experts as the query. \underline{M3} involves enhancing the week expert by incorporating holiday factors. \underline{M4} represents the proposed method, which add the pair-wise combinations of four experts as the
query. The figure demonstrates that incorporating the basic four experts significantly boosts CTCVR performance. Adding the holiday factor improves all metrics, while integrating different expert combinations results in a slight decline in CTCVR GAUC but improves all other metrics.

\textbf{Cold Start Results Comparison.} This section assesses various methods during the cold start phase, focusing on scenarios where user behavior sequences are highly sparse (sequence lengths under 10 in the T+1 day test set). The results, presented in Table 3, clearly show that the proposed method consistently delivers top performance, while other methods experience significant declines. STIM maintained excellent performance across all four metrics, notably improving CTCVR AUC and GAUC by 2.816\% and 2.533\% over the second-best model. These findings highlight STIM's robustness to sparsity in long sequences, as the dynamic masking prioritizes decay from the user's recent behaviors, effectively masking the filler parts of sequences lacking behaviors.

\subsection{Online Experiments}

We assess the model's performance using a primary metric, GTV, which stands for ``Gross Transaction Volume'' and represents the total amount of transactions paid by consumers for products and services on the platform. For this key metric, the proposed STIM demonstrates a significant enhancement of +1.54\% across all users, with statistical significance. This results in daily benefits worth tens of millions for the platform.

\section{Conclusion}
In conclusion, the proposed STIM framework effectively tackles the challenges of long-sequence sparsity and spatiotemporal dependence in local-life service recommendations. By leveraging the forgetting curve for dynamic masking, Query MoE for adaptive expert activation, and a hierarchical network for multi-interest extraction, it captures both recent trends and periodic patterns. This enhances ranking accuracy, offering a robust solution for large-scale local-life recommendation systems.

\bibliography{aaai2026}
\newpage

\begin{figure}

\centering
\includegraphics[width=1\columnwidth]{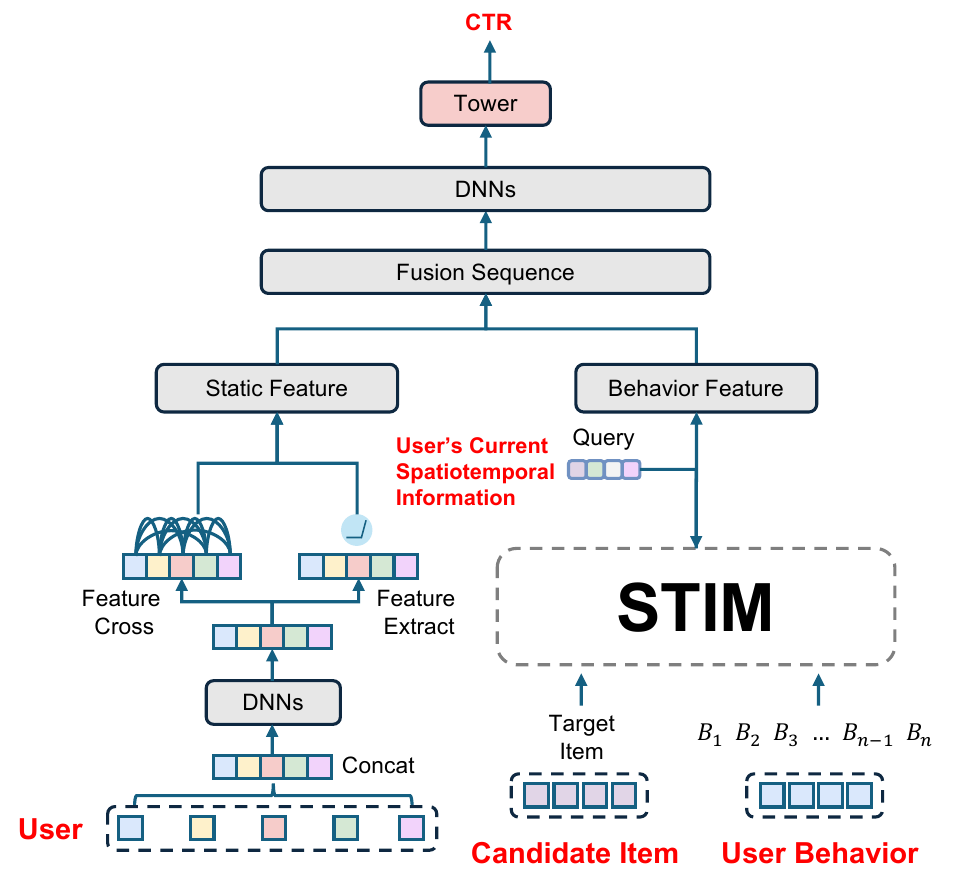} 
\caption{Recommendation pipeline on the Ele.me dataset.}
\label{fig: figS1}
\end{figure}

\begin{figure}

\centering
\includegraphics[width=1\columnwidth]{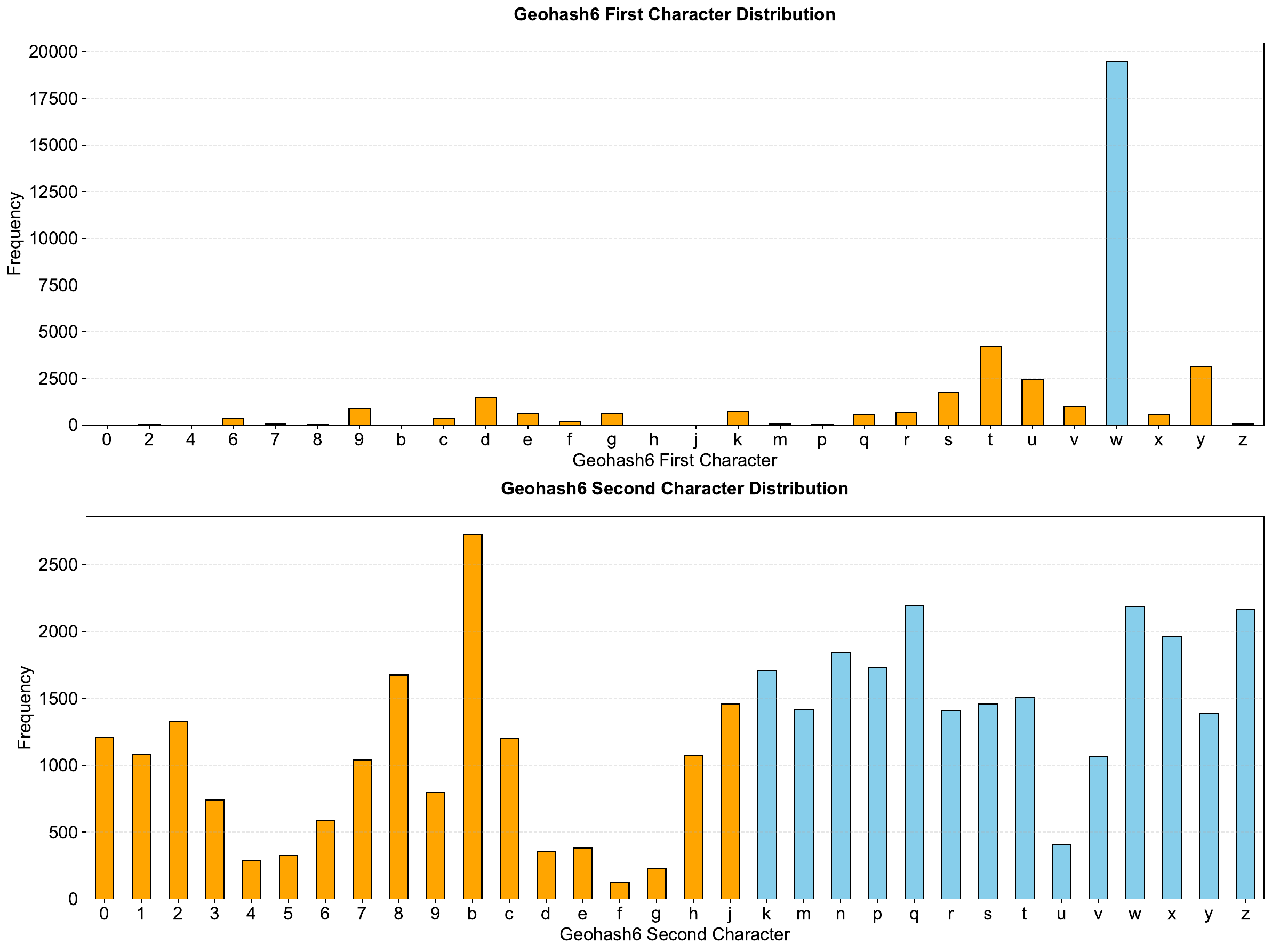} 
\caption{Graph of the frequency distribution of the first and second letters in geohash.}
\label{fig: figS2}
\end{figure}
\section{Entire Recommendation Pipeline under Ele.me Data}
The feature categories and descriptions for the Ele.me dataset are presented in Table \ref{tab:1}. Based on the times found in the User's Current Spatiotemporal Information, one can infer whether the current day is a holiday. Geohash12 is truncated to geohash6. The detailed recommendation pipeline is illustrated in Figure \ref{fig: figS1}. In the figure, user features are processed separately, while candidate items serve as target items. The user behavior and the User's Current Spatiotemporal Information together form the long-sequence behavior modeling component.

\section{Training and Network Details}
For the Ele.me dataset, the STIM model is trained on a single node with 8 V100 GPUs, using a batch size of 100,000 and a learning rate of 1e-4, with the AdamW optimizer for network optimization. As for the TRec dataset, the STIM model is trained across 3 nodes equipped with 24 A100 GPUs, utilizing a batch size of 4,800 and a learning rate of 5e-5, employing the Adam optimizer to achieve optimal performance. For fair comparison, all baseline methods are trained under the identical hardware and hyperparameter configuration as STIM, with all components fixed except for the behavior modeling module. 

\begin{table*}
\renewcommand{\thetable}{S\arabic{table}}
\centering
\begin{tabular}{@{} m{3cm} m{14cm} @{}} 
\toprule
\textbf{Feature Category} & \textbf{Feature Description} \\ \midrule
User & user\_id, gender, visit\_city, avg\_price, is\_supervip, ctr\_30, ord\_30, total\_amt\_30 \\ 

Candidate Item & shop\_id, item\_id, city\_id, district\_id, shop\_aoi\_id, shop\_geohash\_6, shop\_geohash\_12, brand\_id, category\_1\_id, merge\_standard\_food\_id, rank\_7, rank\_30, rank\_90 \\ 

User Behavior & shop\_id\_list, item\_id\_list, category\_1\_id\_list, merge\_standard\_food\_id\_list, brand\_id\_list, price\_list, shop\_aoi\_id\_list, shop\_geohash6\_list, timediff\_list, hours\_list, time\_type\_list, weekdays\_list \\ 
User's Current Spatiotemporal Information & times, hours, time\_type, weekdays, geohash12 \\ 
\bottomrule
\end{tabular}
\caption{Feature Categories and Descriptions}
\label{tab:1}
\end{table*}

The DNN (Deep Neural Network) architecture in Query MoE is a modular, multi-layered neural network designed for customizable feature transformation. Comprising two hidden layers with 8 and 4 neurons respectively, the network dynamically infers input dimensions during initialization. Each hidden layer—except the output layer—applies a configurable activation function (defaulting to ReLU), while the output layer uses a task-specific activation (e.g., softmax for classification).

\section{Different Functional Forms of the Forgetting Curve}
As shown in Figure 4 in the main text, we explored various decay forms of the forgetting curve, including the exponential function used in this paper, as well as power functions and logarithmic functions. The forms of the power function and logarithmic function are respectively: $R(t) = (1 + kt)^m$ and $R(t) = a - b \cdot \ln(t + c)$.

\section{Grouping Statistics Details}

For grouping, time is divided into morning, midday, and evening periods, as well as into days of the week, based on prior knowledge. Geographic location is divided according to the frequency statistics. As shown in Figure \ref{fig: figS2}, this paper statistically analyzes the frequency of the first and second letters in geohash6 for each click. Based on this, geohash is divided into three groups.

\section{Further Demonstrations of Forgetting Curve}

\begin{figure}[h]

\centering
\includegraphics[width=1\columnwidth]{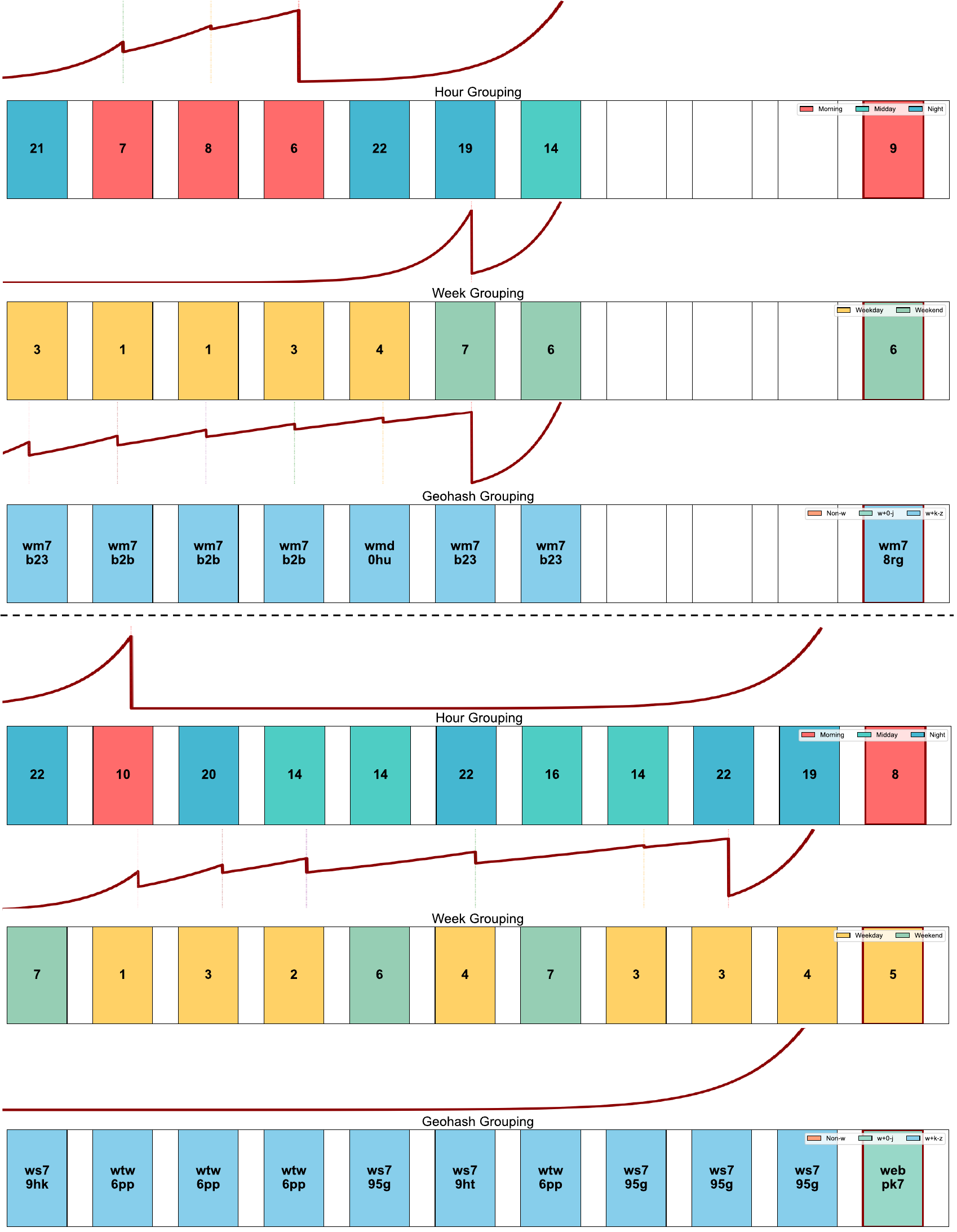} 
\caption{Illustration of the forgetting curve decay from different perspectives in recommendation scenarios.}
\label{figS2}
\end{figure}

More forgetting curve decay graphs are provided, including new scenarios such as those without padding and cases where consecutive geohashes fall within the same group, as shown in Figure \ref{figS2}.

\section{Details for Baselines}
\subsection*{Interest Mining Methods}
\noindent \textbf{DIN (Zhou et al. 2018):} A deep interest network that models user interests through attention mechanisms, focusing on the relevance between target items and historical behaviors. It uses local activation units to capture dynamic interest patterns but does not explicitly handle long-sequence sparsity or spatiotemporal dependencies.

\noindent \textbf{DIEN (Zhou et al. 2019):} Extends DIN by introducing sequential modeling with GRUs and interest evolving units, capturing temporal dependencies in user behavior sequences. It enhances interest representation by distinguishing between positive and negative feedback but lacks dedicated handling of periodic spatiotemporal features.

\subsection*{Long-Sequence Modeling Methods}
\noindent \textbf{SIM (Pi et al. 2020):} A two-stage framework for long-sequence modeling that compresses sequences using GSU (Global Soft Search Unit) to reduce computational complexity. It focuses on extracting global semantic features but does not integrate forgetting mechanisms or spatiotemporal grouping.

\noindent \textbf{TWIN (Chang et al. 2023):} Utilizes a two-tower structure with memory and interaction networks to model long-term and short-term interests separately. It employs hierarchical compression to handle long sequences but prioritizes scalability over fine-grained spatiotemporal pattern mining.

\noindent \textbf{TWIN V2 (Si et al. 2024):} Improves TWIN with hierarchical clustering to compress long sequences more efficiently, enhancing the capture of periodic behaviors. However, it does not incorporate dynamic masking or psychological principles like the forgetting curve.

\subsection*{Temporal Interest Modeling Methods}
\noindent \textbf{IC (Interest Clock, Zhu et al. 2024):} Models user interests with personalized hour-level time encoding, capturing short-term temporal patterns (e.g., daily consumption rhythms). It lacks the ability to model long-term periodicity or spatial factors.

\noindent \textbf{LIC (Long-term Interest Clock, Zhu et al. 2025):} Extends IC by introducing time-gap-aware attention to capture long-term temporal dependencies. It enhances temporal perception but relies on indirect domain transformations (e.g., graph-based methods) for spatiotemporal modeling, without explicit handling of spatial grouping or forgetting mechanisms.

\subsection*{Spatiotemporal Modeling Methods}
\noindent \textbf{FIM (Frequency-Aware Multi-View Interest Modeling, Wang et al. 2025):} A frequency-aware framework designed to capture dynamic periodic patterns in local-life service recommendations. It introduces a multi-view interest decomposition mechanism to distinguish diverse periodicities in user demands, enabling adaptation to scenarios like holidays or promotions. However, it relies on frequency-domain transformations for periodicity modeling and lacks explicit integration of forgetting curves or hierarchical spatiotemporal interaction modeling.

\noindent \textbf{ST-PIL (Spatial-Temporal Periodic Interest Learning, Cui et al. 2021):} Focuses on next point-of-interest recommendation by learning periodic spatiotemporal interests. It incorporates long-term and short-term modules: the long-term module captures daily periodic patterns via day-level granularity, while the short-term module extracts hour-level and region-based spatial-temporal correlations. Despite modeling spatiotemporal periodicity, it lacks adaptive mechanisms for handling sparse long sequences and does not leverage psychological principles like the forgetting curve for dynamic interest weighting.

\begin{table}
\renewcommand{\thetable}{S\arabic{table}}
    \centering
    \begin{adjustbox}{max width=\textwidth}
    \begin{tabular}{ccccc}
    \toprule
    \multirowcell{2}{\textbf{Modules}} & \multicolumn{2}{c}{\textbf{CTR}} & \multicolumn{2}{c}{\textbf{CTCVR}} \\
    \cmidrule(lr){2-3} \cmidrule(lr){4-5}
    & \textbf{AUC} & \textbf{GAUC} & \textbf{AUC} & \textbf{GAUC} \\
    \midrule
    MHA & 0.7674 & 0.6911 & 0.8698 & 0.7547 \\
    HMIN-Unit & 0.7674 & 0.6916 & 0.8710 & 0.7551 \\
    \bottomrule
    \end{tabular}
    \end{adjustbox}
    \caption{Performance comparison of Multi-Head Attention and HMIN-Unit.}
    \label{tab:overall_small}
\end{table}

\section{Ablation experiment of the HMIN-Unit}
In this section, as shown in Table \ref{tab:overall_small}, we compared the HMIN-Unit with Multi-Head Attention (MHA) in terms of CTR AUC and GAUC, as well as CTCVR AUC and GAUC. The results indicate that the HMIN-Unit outperforms MHA significantly in CTR GAUC, CTCVR AUC, and GAUC, with improvements of 0.072\%, 0.138\%, and 0.053\%, respectively. This demonstrates that the HMIN-Unit, while being more lightweight, effectively accomplishes feature interactions across different levels and facilitates multi-interest exploration.

\end{document}